\documentclass[letterpaper]{article} 
\usepackage{aaai24}  
\usepackage{times}  
\usepackage{helvet}  
\usepackage{courier}  
\usepackage[hyphens]{url}  
\usepackage{graphicx} 
\usepackage{natbib}  
\usepackage{caption} 
\usepackage{algorithm}
\usepackage{algorithmic}

\usepackage{newfloat}
\usepackage{listings}
\DeclareCaptionStyle{ruled}{labelfont=normalfont,labelsep=colon,strut=off} 
\floatstyle{ruled}
\newfloat{listing}{tb}{lst}{}
\floatname{listing}{Listing}
\usepackage{subcaption}
\usepackage{graphicx}
\usepackage{amsmath}
\usepackage{amsfonts}
\usepackage{bm}
\usepackage{xspace}
\usepackage{tabularx, multirow, multicol}
\usepackage{color, colortbl}
\usepackage{xcolor}
\usepackage{booktabs}
\usepackage{multirow}
\usepackage{enumitem}
\usepackage{tabularx, booktabs}
\usepackage{xspace}
\newtheorem{definition}{Definition}

\newcommand{\msa}{\textnormal{MSA}\xspace}
\newcommand{\satt}{\textnormal{SAtt}\xspace}
\newcommand{\satth}{\textnormal{SAttH}\xspace}

\newcommand{\stoken}{\texttt{[S]}\xspace}
\newcommand{\mtoken}{\texttt{[M]}\xspace}
\newcommand{\stokenemb}{\bm{e}_{\texttt{[S]}}}
\newcommand{\mtokenemb}{\bm{e}_{\texttt{[M]}}}

\newcommand{\smallsection}[1]{{\noindent \bf {#1.\hspace{5pt}}}}

\newcommand{\scenario}{MDRAU\xspace}
\newcommand{\proposed}{\textnormal{DRIP}\xspace}

\newcolumntype{P}{>{\centering\arraybackslash}m{0.07\linewidth}}
\definecolor{Gray}{gray}{0.9}

\def\R{{\bm{R}}}
\def\Rk{{\bm{R}^{(k)}}}
\def\U{{\mathcal{U}}}
\def\Uk{{\mathcal{U}_k}}
\def\Vk{{\mathcal{V}_k}}
\def\V{{\mathcal{V}}}
\def\D{{\mathcal{D}}}

\title{Multi-Domain Recommendation to Attract Users via Domain Preference Modeling}
\author{
    Hyunjun Ju\textsuperscript{\rm 1}, 
    SeongKu Kang\textsuperscript{\rm 2}, 
    Dongha Lee\textsuperscript{\rm 3}, 
    Junyoung Hwang\textsuperscript{\rm 1}, 
    Sanghwan Jang\textsuperscript{\rm 1}, 
    Hwanjo Yu\textsuperscript{\rm 1}\thanks{Corresponding Author} \\
}
\affiliations{
    \textsuperscript{\rm 1} Pohang University of Science and Technology (POSTECH), Republic of Korea\\
    \textsuperscript{\rm 2} University of Illinois at Urban-Champaign (UIUC), United States \\
    \textsuperscript{\rm 3} Yonsei University, Republic of Korea\\

    \{hyunjunju, jyhwang, s.jang, hwanjoyu\}@postech.ac.kr, seongku@illinois.edu, donalee@yonsei.ac.kr
}

\begin{document}

\maketitle

\begin{abstract}
    Recently, web platforms have been operating various service domains simultaneously.
Targeting a platform that operates multiple service domains, we introduce \textit{a new task}, Multi-Domain Recommendation to Attract Users (MDRAU), which recommends items from multiple ``unseen'' domains with which each user has not interacted yet, by using knowledge from the user's ``seen'' domains.
In this paper, we point out two challenges of MDRAU task.
First, there are numerous possible combinations of mappings from seen to unseen domains because users have usually interacted with a different subset of service domains.
Second, a user might have different preferences for each of the target unseen domains, which requires that recommendations reflect the user's preferences on domains as well as items.
To tackle these challenges, we propose \proposed framework that models users' preferences at two levels (i.e., domain and item) and learns various seen-unseen domain mappings in a unified way with masked domain modeling.
Our extensive experiments demonstrate the effectiveness of \proposed in MDRAU task and its ability to capture users' domain-level preferences.
\end{abstract}

\section{Introduction}
Nowadays, web platforms are operating various service domains simultaneously (e.g., music streaming, game store, and eBook subscription).
They allow users to experience diverse domains within a single platform and promote the mutual growth of all service domains through the Recommender System (RS).
For such multi-domain platforms, recommending items from \textit{unseen domains} with which each user has not interacted yet plays an essential role in the platform's growth, user satisfaction, and business success.
That is, users typically utilize a few domains rather than all domains, and accurate recommendations that align with user preference can attract users into unexplored domains.
To this end, Cross-Domain Recommendation (CDR), which recommends items from unseen (target) domains based on user interaction history in seen (source) domains, has gained significant research attention.\footnote{In this paper, the terms ``seen'' and ``unseen'' are defined from the perspective of each user;
as each user interacts with a different subset of service domains, the seen and unseen domain sets vary for each user.
A user can be considered a cold-start user in the user's unseen service domain~\cite{emcdr,sscdr, ptupcdr}.}

\begin{figure}[t]
    \centering
    \includegraphics[width=0.83\columnwidth]{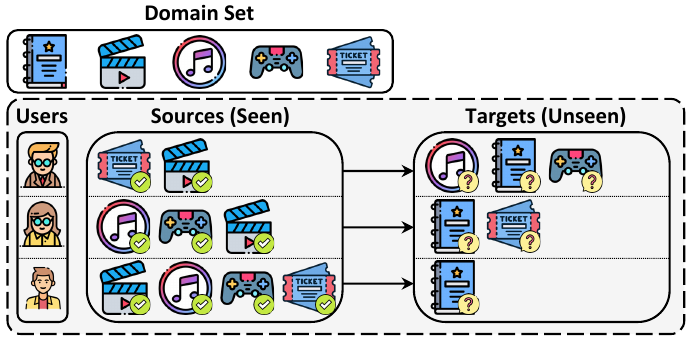}
    \caption{A conceptual illustration of MDRAU task. 
    A platform operates five different service domains, and each user partially interacts with a subset of the entire service domains.
    MDRAU aims to provide recommendations from each user's unseen domains to attract users.}
    \label{fig:figure1}
\end{figure}

Most CDR studies~\cite{emcdr, sscdr, ptupcdr, tmcdr, deeplyfusing} have focused on transferring user preference information from a source domain to a target domain.
Given a recommender system employed for each domain, they learn a mapping function that acts as a bridge between the representation spaces of the two domains.
To provide unseen domain recommendations, they transfer the user embedding from the source domain to the target domain using the mapping function and generate recommendations based on the transferred user embedding.
More recently, \cite{cdrib, unicdr, disencdr} have achieved improved recommendation accuracy by modeling domain-specific and domain-shared information separately, and selectively transferring the domain-shared information.
Despite their effectiveness, the existing studies have targeted the case of a single seen-unseen domain pair (i.e., one-to-one), and the case of multiple seen-unseen domains (i.e., many-to-many) has not been studied well.
Especially, they do not consider making unified recommendations that include items from multiple domains.

In practical scenarios, it is becoming increasingly common for platforms to offer services in more than two domains.
This creates a need to promote user engagement across multiple unseen domains by attracting users with accurate personalized recommendations.
We refer to this problem as Multi-Domain Recommendation to Attract Users (MDRAU) task (Fig.~\ref{fig:figure1}).
Formally, MDRAU aims to provide a recommendation list that consists of items from each user's unseen domain(s) that the user has not tried before.
MDRAU brings several practical values to multi-domain platforms.
It encourages users to explore new content beyond their previously interacted domains, fostering diverse and serendipitous discoveries.
This diversified user experience can lead to enhanced user satisfaction and engagement, which in turn helps to provide more accurate recommendations in each domain.

Addressing MDRAU task presents two major challenges.
First, since each user has interacted with different service domains, there are numerous possible combinations of seen-unseen domains.
With $K$ domains, the number of combinations can reach up to $2^K-2$, except in the case that a user has used all services.
This large number of combinations makes it difficult to apply the previous CDR methods that learn a one-to-one mapping function for each domain pair.
Second, a user naturally has different preference for each unseen domain, and these varying domain preference need to be properly reflected in the recommendation process.
That is, the recommendation needs to consider user preference in both \textit{domain-level}, i.e., the inclination of a user to explore each unseen domain, and \textit{item-level}, i.e., the inclination of a user to interact with a new item within a domain.
The previous CDR methods have focused on improving item-level preference for a specific unseen domain without directly considering domain-level preference.
As a result, they show limited performance when applied to \scenario task.

To effectively solve our new task MDRAU, we propose \proposed, a new framework that learns various seen-unseen domain mappings in a unified way via masked domain modeling and models user preference at the domain and item levels.
We formulate the training process of \proposed as a prediction task of missing information based on its contexts \cite{BERT, beit}.
Then, we model the two-level preferences using a multi-domain encoder that incorporates user preference across multiple domains.
The key idea is to randomly mask the user preference of some seen domains in the model input, and train the model to predict the user preference in the masked domains.
During the training, we regard the masked domains as the user's unseen domains, allowing the model to simulate and learn from \textit{numerous scenarios} involving different combinations of seen and unseen domains.
This enables the model to achieve the generalization capability of inferring user preferences in unseen domains from ones in seen domains.
Furthermore, we introduce an adaptive masking scheme to make the model more focused on learning domains that a user is more likely to prefer.
We validate the superiority of \proposed by extensive experiments on real-world datasets and provide a thorough comparison with various state-of-the-art methods.

\section{Related Work}
The existing CDR studies can be divided into two groups according to the type of target domain for recommendations.

\smallsection{CDR for Seen Domain Recommendation}
It aims to improve the recommendation quality of seen domains with which the user has already interacted.
Many studies alleviate the data sparsity problem in the sparse target domains by utilizing information from the source domain~\cite{conet, dtcdr, darec, GA-dtcdr, bitgcf, ddtcdr, disencdr}.
To this end, they transfer knowledge among domains via bridging information, such as overlapping users or items.
For example, CoNet~\cite{conet} introduces cross-connection units to transfer and integrate knowledge between source and target domains.
DTCDR~\cite{dtcdr} proposes a dual-target framework to improve the recommendation accuracy in both two involved domains simultaneously.
GA-DTCDR~\cite{GA-dtcdr} extends DTCDR by adopting graph information. 
Recently, several studies have focused on multi-domain cases having more than two domains~\cite{kdd08, mcf,unicdr, cat_art, msdcr, cui2020herograph, ga_mtcdr}.
GA-MTCDR~\cite{ga_mtcdr} extends GA-DTCDR with element-wise attention to integrating embeddings of overlapping users from multiple domains.
CAT-ART~\cite{cat_art} proposes a contrastive autoencoder to encode a global user embedding and a mechanism to transfer user embeddings from each source domain to the target domain.
UniCDR~\cite{unicdr} introduces domain-specific and domain-shared embeddings along with aggregation schemes to make a universal model for existing CDR scenarios.

\smallsection{CDR for Unseen Domain Recommendation}
It aims to provide recommendations in unseen domains with which the user has not yet interacted.
Their focus lies on the method of obtaining user embeddings in the target domain space.
For example, EMCDR~\cite{emcdr} proposes an embedding and mapping framework, which learns a mapping function that transfers user embeddings from the source to the unseen target domain.
SSCDR~\cite{sscdr} proposes a semi-supervised embedding and mapping framework to train a mapping function, even when only a few labeled data are available.
PTUPCDR~\cite{ptupcdr} uses a meta-network that generates a personalized mapping function.
UniCDR~\cite{unicdr} can also be applied to recommend unseen domains using domain-shared embeddings.
They mainly focus on a single unseen target domain rather than multiple unseen target domains.

\smallskip
\noindent
Despite their effectiveness, there is no method particularly designed for MDRAU task. 
Existing CDR methods that handle unseen domains have only focused on the case of a single seen-unseen domain pair. 
Also, CDR methods dealing with multi-domain cases mainly focus on recommendations within seen domains.
To address MDRAU task, a method needs the ability to create an accurate, unified recommendation list across multiple domains that users have not tried before.
We believe that a tailored solution for MDRAU task is required.
\section{Problem Formulation}
\label{sec:formulation}
\subsection{Notations}
In this work, we focus on a scenario where a provider operates services for multiple domains (e.g., music streaming, game store, and eBook subscription), each of which employs a distinct recommender system.
Each service domain has its own user and item set. 
Items of each domain are mutually exclusive, while users may use one or multiple service domain(s).
Formally, given $K$ domains $\mathcal{D}=\{d_1,\cdots,d_K\}$, $\Uk$ and $\Vk$ denote the set of users and items for the $k$-th domain, respectively.
The user-item interaction history for $d_k$ is represented by a matrix $\Rk\in \left\{0, 1\right\}^{\vert\Uk\vert\times\vert\Vk\vert}$, where $R_{u,v}=1$ if user $u$ has interacted with the item $v$, otherwise $R_{u,v}=0$.
Without loss of generality, we define an interaction matrix of all domains $\R\in\{0, 1\}^{\vert\U\vert\times \vert\V\vert}$, where $\U=\bigcup_{k=1}^{K}\Uk$ and $\V=\bigcup_{k=1}^{K}\Vk$.
We additionally define the user-domain relations as $\bm{G}\in \left\{0, 1\right\}^{|\mathcal{U}|\times |\mathcal{D}|}$, where $G_{u, k}=1$ if the user $u$ has interacted with items of domain $d_k$, otherwise $G_{u,k}=0$. 
Overlapping users indicate users who have interacted with at least two domains.

\subsection{MDRAU Task}
\begin{definition}[\small Multi-Domain Recommendation to Attract Users]
Given user-item interaction history from multiple service domains, MDRAU refers to the task of providing a ranking list (i.e., recommendation) that consists of items from each user's unseen domain(s) that the user has not tried before.
\end{definition}

\noindent
In some cases, providers may have a need to promote a specific target domain within the platform.
We refer to the scenario with a single target domain as MDRAU-ST, and the scenario with multiple target domains as MDRAU-MT.
MDRAU provides preferable items from each user's unseen domains, encouraging the exploration of new content beyond their previously interacted domains.
By doing so, MDRAU helps to diversify user experience and facilitate serendipitous discoveries, enhancing user engagement.
This enhanced engagement, in turn, helps to provide more accurate recommendations in each domain, ultimately contributing to the growth and revenue of the platform.
\begin{figure}[t]
    \centering
    \includegraphics[width=0.95\columnwidth]{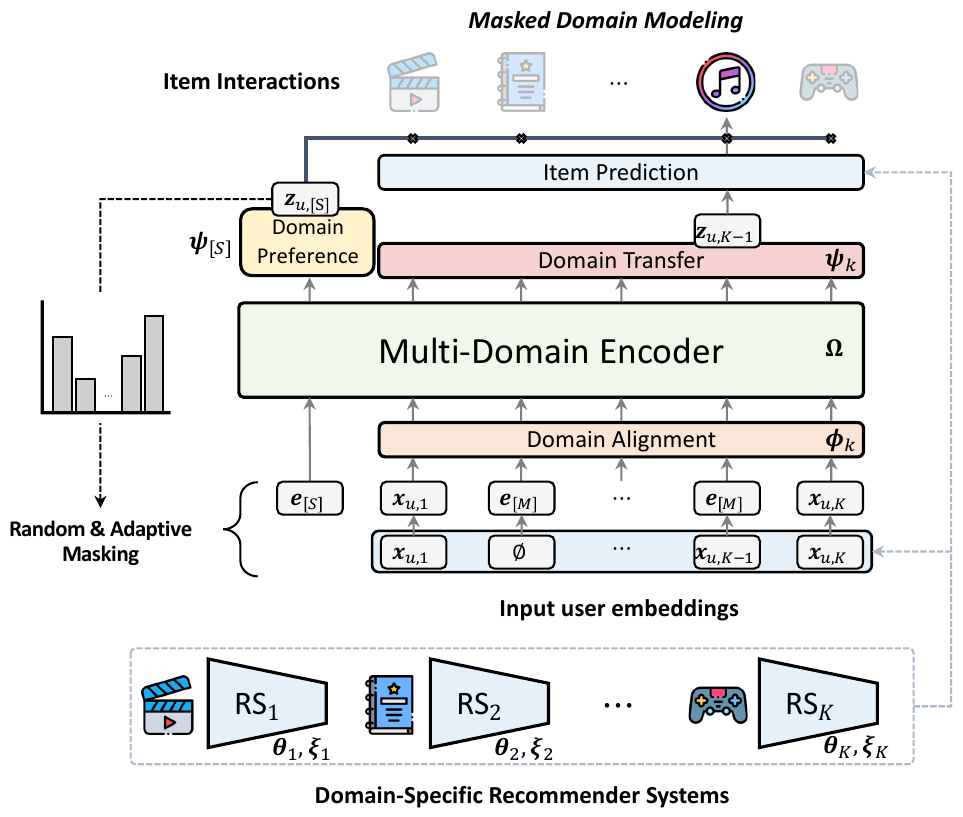} 
    \caption{The overview of the proposed \proposed framework.} 
    \label{fig:overall}
\end{figure}
\section{Proposed Framework}
\label{sec:proposed}

\subsection{Overview}
\label{subsec:overview}
We present a unified framework for MDRAU task, named as \proposed (\underline{D}omain p\underline{R}eference-aware unseen domain \underline{I}tem \underline{P}rediction).
\proposed optimizes the parameters of its model by maximizing the likelihood given the training data $\R$.
Let $\bm{R}_u \in \{0, 1\}^{\vert\mathcal{V}\vert}$ denote a multi-hot vector representing a user's interaction with the items over all domains.
We assume the observed data are drawn from the Multinomial distribution, and the likelihood is described by
\begin{equation}
    \begin{split}
        p(\bm{R}_u)&=\prod_{v\in\mathcal{V}}p(v|u)^{R_{u,v}} \\
        \bm{R}_u &\sim \textnormal{Mult}(N_u, p(v|u)),
    \end{split}
\end{equation}
where $N_u=\sum_v R_{u,v}$ and $p(v|u)$ is the probability that user $u$ prefers the item $v$ over the entire item set.
We decompose the likelihood based on the domain-level and item-level preference as follows:
\begin{subequations}
\label{eq:likelihood}
\begin{align}
    p(\bm{R}_u) &=\prod_{v\in\V}p(v|u)^{R_{u,v}} 
                = \prod_{d_k\in\D}\prod_{v\in\V_k} p(v, d_k|u)^{R_{u,v}} \label{eq:line2}\\
                &= \prod_{d_k\in\D}\prod_{v\in\V_k} \big[p(v|u, d_k)\cdot p(d_k|u)\big]^{R_{u,v}},
\end{align}    
\end{subequations}
where $p(v|u, d_k)$ denotes user $u$'s preference for item $v$ in domain $d_k$ and $p(d_k|u)$ denotes the user's preference for domain $d_k$.
Eq.~\eqref{eq:line2} holds because item sets from different domains are mutually exclusive from each other.
These preferences are modeled by a unified neural model with a multi-domain encoder based on self-attention.
Then, to maximize the likelihood, we train the model via \textit{masked domain modeling} that predicts the item preference of the masked domains.
The recommendations are produced by considering both domain- and item-level preferences for unseen domains of each user.
Fig.~\ref{fig:overall} illustrates the overall \proposed architecture.

\subsection{Domain-Specific Encoder}
\label{subsec:encoder}
As aforementioned, we assume that a service platform operates multiple service domains and has deployed its own RS for each domain.
An RS model typically contains \textit{encoders} that encode the user and item information into representation space, where the user-item similarity is measured for recommendations.
In specific, for each domain $d_k$, let $f_{\bm{\theta}_k}: \mathbb{R}^{\vert\mathcal{U}_k\vert}\rightarrow\mathbb{R}^d$, $f_{\bm{\xi}_k}:\mathbb{R}^{\vert\mathcal{V}_k\vert} \rightarrow\mathbb{R}^d $, and $\text{sim}_k(\cdot, \cdot)$ denote a user encoder, an item encoder, and similarity function, respectively.
A variety of architectures can be adopted for the encoders (e.g., id-based \cite{koren_mf, bpr} and graph-based~\cite{ngcf, lgcn}), and $\text{sim}_k(\cdot, \cdot)$ can be either a simple metric or a learnable function.
In this work, we use a simple id-based encoder with the inner product similarity, as done in~\cite{bpr, emcdr, ptupcdr}.

\subsection{Multi-Domain Encoder}
\label{subsec:multi-domain-encoder}
The main component of our unified neural model is a \textit{multi-domain encoder} that aims to enrich user embedding from each domain-specific encoder with that from the encoders for other domains.
It adopts the self-attention mechanism to aggregate information from other domains based on the similarity of user preferences across multiple domains.

\subsubsection{Constructing Masked Input.}
\label{subsubsec:construct}
Let $\bm{x}_{u,k}=f_{\bm{\theta}_k}(u)$ denote user embedding of user $u$ in domain $d_k$.
A user $u$ is represented as the set of the corresponding user embeddings for all domains $\{\bm{x}_{u,k}\}_{k=1}^K$. 
Note that some domains may have no interaction history with the user (i.e., $G_{u,k}=0$), as users typically utilize a few domains rather than all domains. 
To handle this case, we replace the embedding for the user's unseen domains with $\mtokenemb$, which is the learnable embedding of a special mask token \mtoken as follows:
\begin{equation}
\label{eq:input_emb}
    \bar{\bm{X}}_u=\{\bar{\bm{x}}_{u,k}\}_{k=1}^K, \quad \bar{\bm{x}}_{u,k}=(1-G_{u, k}) \mtokenemb + G_{u, k}\bm{x}_{u,k}.
\end{equation}

Because the embeddings are obtained from independently trained domain-specific RS models, they have different distributions. 
We align the distributions using projectors $g_{\bm{\phi}_k}(\bm{x}_{u,k})$, where $g_{\bm{\phi}_k}: \mathbb{R}^d \rightarrow \mathbb{R}^m$.
Besides, we insert $\bm{e}_{\texttt{[S]}}\in\mathbb{R}^m$ at the beginning of the input for the encoder, which is the learnable embedding of a special token \texttt{[S]};
this will be used to estimate each user's domain-level preference.
The final input representation of user $u$ is constructed as 
\begin{equation}
\label{eq:input_rep}
        \bm{H}_u^0=\big[\stokenemb, g_{\bm{\phi}_1}(\bar{\bm{x}}_{u, 1}), \cdots, g_{\bm{\phi}_K}(\bar{\bm{x}}_{u,K})\big]^{\top} \in \mathbb{R}^{(K+1) \times m}.
\end{equation}
We do not use position embedding because spatial position information is unnecessary for our target task.

\subsubsection{Contextualizing User Embeddings over Multi-Domains.}
The input $\bm{H}_u^0$ is forwarded into the multi-domain encoder to contextualize each domain-specific user embedding over the user's multiple domains based on the self-attention mechanism.
The multi-domain encoder is basically a stack of $L$ transformer layers \cite{transformer, BERT}.
The details are described in the Appendix.
The $l$-th transformer layer can be simply described by 
\begin{equation}
    \bm{H}_u^{l+1} = \text{Transformer}(\bm{H}_u^{l}), \quad \forall l\in \left\{0, 1, \cdots, L-1\right\}.
\end{equation}
We denote a set of learnable parameters in the multi-domain encoder consisting of transformer layers as $\bm{\Omega}$ for a consistent explanation.
In the end, the final output of $L$-th layer is obtained by 
\begin{equation}
    \bm{H}_u^L=\big[\bm{h}_{u, \texttt{[S]}}^L, \bm{h}_{u, 1}^L, \cdots, \bm{h}_{u,K}^L\big]^{\top} \in \mathbb{R}^{(K+1) \times m}.
\end{equation}
Through the transformer layers, the user embedding from each domain gets contextualized by attending to the embeddings from other domains based on the embedding similarity across the domains.
As a result, $\bm{h}_{u, k}^L$ encodes user preference for domain $d_k$, being enriched by preference information from the other remaining domains.

\subsection{Preference Modeling}
\label{subsec:preference-modeling}
\subsubsection{Domain-level preference.}
The domain-level preference can be inferred by using the contextualized representation of the special token \stoken for user $u$, denoted by $\bm{h}_{u, \stoken}^L$, which is generated by aggregating a user's preference information for multiple domains.
We introduce a domain-preference predictor $q_{\bm{\psi}_{\stoken}}: \mathbb{R}^m \rightarrow \mathbb{R}^K$ to predict each user's inclination to each domain: $\bm{z}_{u, \stoken} = q_{\bm{\psi}_{\stoken}}(\bm{h}_{u, \stoken}^L)$.
The domain-level preference is defined by
\begin{equation}
\label{eq:dom-level-pref}
    p(d_k|u; \bm{\Theta}) = \frac{\exp{(\bm{z}_{{u,\stoken}_{k}})}}{\sum_{{d}_i\in\mathcal{D}}\exp{(\bm{z}_{{u,\stoken}_{i}})}},
\end{equation}
where $\bm{z}_{{u, \stoken}_k}$ indicates $k$-th logit value in $\bm{z}_{u, \stoken}$.

\subsubsection{Item-level preference.} 
The item-level preference in each domain is obtained from the similarity of item embeddings to the contextualized user embedding for the domain.
Using a domain-specific projector $q_{\bm{\psi}_k}: \mathbb{R}^m \rightarrow \mathbb{R}^d$, we first project $\bm{h}_{u, k}^L$ to the representation space of domain $d_k$: $\bm{z}_{u,k}=q_{\bm{\psi}_k}(\bm{h}_k^L)$.
We compute the in-domain item-level preference based on user $u$'s similarity distribution over item set $\V_k$,
\begin{equation}
\label{eq:item-level-preference}
    p(v|u,d_k;\bm{\Theta}) = \frac{\exp{\big(\textnormal{sim}_k(\bm{z}_{u,k}, \bm{x}_{v,k})\big)}}{\sum_{\hat{v}\in\V_k}\exp{\big(\textnormal{sim}_k(\bm{z}_{u,k}, \bm{x}_{\hat{v}, k})}\big)},
\end{equation}
where $\bm{x}_{v,k}=f_{\bm{\xi}_k}(v)$ is the embedding of item $v$ in the domain $d_k$.

\subsection{Model Learning}
\label{subsec:learning}
\subsubsection{Masked Domain Modeling}
We formulate the training process of \proposed as a prediction task of missing domain information based on its contexts \cite{BERT, beit}.
Our key idea is to randomly mask some of the domain-specific user embeddings (among the ones for a user's seen domains) in the input, and train the model to predict the user preference in the masked domains.
That is, in the training process, we regard randomly masked domains as the user's unseen domains, which allows our model to simulate and learn various scenarios of mapping user preference from seen domains to unseen domains.
As a result, the model can capture the relations of user preferences across domains, eventually achieving the generalization capability of inferring user preferences in unseen domains from those in seen domains.

Let $\bm{m}_u\in \{0, 1\}^K$ denote a random masking vector for user $u$, where $m_{u,k}=1$ indicates that the user embedding for domain $d_k$ is masked.
We apply the making operation to the user embedding for the seen domains (i.e., $G_{u,k}=1$) with the probability of $p_{u,k}$.
In specific, $m_{u,k}$ is drawn from a Bernoulli distribution: $m_{u,k}\sim \text{Bern}(p_{u,k})$.
At the beginning of training, we set the equal masking probability for all seen domains (i.e., random masking).
Then, during the training process, we gradually adjust the probability based on the domain-level preferences, i.e., $p_{u,k} \propto p(d_k|u;\Theta)$, to encourage the model to more focus on learning domains that the user is more likely to prefer (i.e, adaptive masking).
The masked embedding set is represented as $\Tilde{\bm{X}}_u= \{\Tilde{\bm{x}}_{u,k}\}_{k=1}^K$, where its elements are obtained by
\begin{equation}\label{eq:masking-train}
    \Tilde{\bm{x}}_{u, k} = (1-G_{u,k})\bm{e}_{\texttt{[M]}} +  G_{u,k}\big(m_{u,k} \bm{e}_{\texttt{[M]}} + (1-m_{u,k}) \bm{x}_{u,k}\big).
\end{equation}
During the training, we use $\Tilde{\bm{X}}_u$ (Eq.~\eqref{eq:masking-train}) instead of $\bar{\bm{X}}_u$ (Eq.~\eqref{eq:input_emb}) for the input of the multi-domain encoder, and obtain their contextualized representations.
Note that we discard the case in which all seen domains are masked.

\subsubsection{Learning objective}
We train the model to predict the preference information of the masked domains (i.e., $m_{u,k}=1$).
Instead of predicting the masked embedding itself, we directly maximize the likelihood of the user's interaction history.
This makes the recommendation accuracy directly aligned with the optimization objective.
Based on the negative log-likelihood of Eq.~\eqref{eq:likelihood} and the masking process, the final loss is defined as follows:

\begin{equation}
\label{eq:loss}
    \small\mathcal{L}_{\proposed} = 
                 - \sum_{u\in\mathcal{U}}\sum_{d_k\in \mathcal{D}}m_{u,k} \sum_{v\in\mathcal{V}_k} R_{u,v}\log [p(v|u,d_k;\bm{\Theta})\cdot p(d_k|u;\bm{\Theta})],
\end{equation}
where the learning parameters $\bm{\Theta}$ include the multi-domain encoder $\bm{\Omega}$, two types of projectors $\{\bm{\phi}_k\}_{k=1}^K$ and $\{\bm{\psi}_k\}_{k=1}^K$, and domain-preference predictor $\bm{\psi}_{\stoken}$.

\subsection{MDRAU Recommendation}
At the test phase, for each user, we construct the input representation with the user's unseen domains masked (Eq.~\eqref{eq:input_rep}), and calculate the domain-level and item-level preferences, $p(d_k|u; \bm{\Theta})$ and $p(v|u, d_k;\bm{\Theta})$.
In the scenario where there are multiple target service domains (MDRAU-MT), the recommendation is generated by sorting the items by their score $p(v|u, d_k;\bm{\Theta}) \cdot p(d_k|u; \bm{\Theta})$.
Otherwise, in the scenario where we have a specific target service domain to promote (MDRAU-ST), the recommendation is generated by considering only in-domain item-level preference $p(v|u, d_k;\bm{\Theta})$.

\section{Experiments}
\subsection{Experimental Settings\footnote{More detailed experimental settings are in the Appendix.}} 
\subsubsection{Datasets and Domain Setup}
We use the widely-used Amazon dataset~\cite{amazon, sscdr, cdrsurvey}, which consists of multiple item domains.
To simulate the platform environment, we select two subsets of these domains that have been previously used in related studies~\cite{cdrsurvey}.
The first platform scenario (P1) includes Book, Movie, CD, and Game domains, and the second scenario (P2) includes Home, Health, Grocery, and Tools domains.
The detailed statistics of the datasets are described in Table~\ref{table:statistics} and the Appendix.

\subsubsection{Evaluation Metrics}
We focus on the top-$K$ recommendation task for implicit feedback.
We evaluate the recommendation accuracy of each method by using the two ranking metrics~\cite{cml,vaecf}: Recall (R@$K$) and Normalized Discounted Cumulative Gain (N@$K$).

\subsubsection{Compared Methods}
We compare \proposed with various methods from related research fields.
We have modified the original methods to perform MDRAU, and the modified versions are annotated with the suffix `+'.
The \textit{first group} of baselines learns a recommendation model \textbf{BPRMF}~\cite{bpr} while treating a union of multiple domains as a single domain~\cite{kdd08}.
The \textit{second group} includes multi-task learning methods (\textbf{MMOE}~\cite{mmoe} and \textbf{PLE}~\cite{ple}), which define a single task as item-level preference learning for each domain.
They are widely used for RS in multi-domain cases~\cite{unicdr}.
For each task (i.e., domain), they employ the binary cross-entropy loss to predict user-item interactions from implicit feedback \cite{he2017neural}.
The \textit{third group} includes CDR methods (\textbf{EMCDR+}~\cite{emcdr} and \textbf{PTUPCDR+}~\cite{ptupcdr}) for unseen domain recommendation, which learn a mapping function for each source-target domain pair.
Due to a large number of seen-unseen domain combinations, it is infeasible to directly apply them to MDRAU.
For this reason, we tailor their learning task for a many-to-one mapping.
The \textit{last group} includes state-of-the-art methods (\textbf{CAT-ART+}~\cite{cat_art} and \textbf{UniCDR}~\cite{unicdr}) that partially handle multi-domain cases.
They are designed to exploit information from multiple domains to improve recommendation accuracy in each domain.
For \scenario-MT, recommendations from each unseen domain are integrated into a unified recommendation list using post-processing (e.g., normalization).
\begin{table}[t]
\centering
\resizebox{0.8\columnwidth}{!}{
\begin{tabular}{@{}cccccc@{}}
\toprule
& Domains & \#Users & \#Items & \#Interaction & Density \\ \midrule
\multirow{4}{*}{P1} & Book & 35,987 & 39,049 & 1,726,231 & 0.12\% \\
                           & Movie & 17,056 & 15,620 & 609,552 & 0.23\% \\
                           & CD    & 5,941 & 9,069 & 211,617 & 0.39\% \\
                           & Game  & 1,049 & 1,064 & 20,568 & 1.84\% \\ \midrule
\multirow{4}{*}{P2} & Home & 10,317 & 7,605 & 127,091 & 0.16\% \\
                           & Health & 8,690 & 5,750 & 123,905 & 0.25\% \\
                           & Grocery & 4,869 & 2,884 & 71,269 & 0.51\% \\
                           & Tools & 1,804 & 1,422 & 18,040 & 0.70\% \\ \bottomrule
\end{tabular}}
\caption{Statistics of the two platform scenarios of \scenario.}
\label{table:statistics}
\end{table}
\begin{table*}[thbp]
\centering
\resizebox{\textwidth}{!}{%
\begin{tabular}{@{}ccccccccccccccccc@{}}
\toprule
\multirow{2}{*}{\textbf{Methods (P1)}}    & \multicolumn{4}{c}{\textbf{Book}}  & \multicolumn{4}{c}{\textbf{Movie}} & \multicolumn{4}{c}{\textbf{CD}} & \multicolumn{4}{c}{\textbf{Game}} \\  \cmidrule(l){2-5} \cmidrule(l){6-9} \cmidrule(l){10-13} \cmidrule(l){14-17}
  & \textbf{R@20} & \textbf{R@50} & \textbf{N@20} & \textbf{N@50} & \textbf{R@20}  & \textbf{R@50}  & \textbf{N@20} & \textbf{N@50} & \textbf{R@20} & \textbf{R@50} & \textbf{N@20} & \textbf{N@50} & \textbf{R@20}  & \textbf{R@50} & \textbf{N@20} & \textbf{N@50} \\ \midrule
BPRMF & 0.0174  & 0.0176  & 0.0182  & 0.0205  & 0.0223  & 0.0357  & 0.0224  & 0.0279  & 0.0378  & 0.0613  & 0.0371  & 0.0471  & 0.0572  & 0.1315  & 0.0548  & 0.0868 \\ \midrule
MMOE & 0.0161  & 0.0183  & 0.0164  & 0.0171  & 0.0306  & 0.0423  & 0.0310  & 0.0352  & 0.0338  & 0.0535  & 0.0341  & 0.0408  & 0.0729  & 0.1467  & 0.0613  & 0.0941 \\
PLE & 0.0103  & 0.0149  & 0.0108  & 0.0123  & 0.0307  & 0.0469  & 0.0296  & 0.0368  & 0.0375  & 0.0499  & 0.0337  & 0.0382  & 0.0750  & 0.1379  & 0.0592  & 0.0872 \\ \midrule
EMCDR+ & 0.0322  & 0.0460  & 0.0348  & 0.0396  & 0.0449  & 0.0647  & 0.0439  & 0.0523  & 0.0574  & 0.0767  & 0.0610  & 0.0673  & 0.0846  & 0.1600  & 0.0755  & 0.1093 \\
PTUPCDR+ & 0.0311  & 0.0417  & 0.0338  & 0.0373  & 0.0426  & 0.0642  & 0.0427  & 0.0516  & 0.0559  & 0.0778  & 0.0589  & 0.0669  & 0.0875  & 0.1535  & 0.0836  & 0.1131 \\ \midrule
CAT-ART+ & 0.0325  & 0.0426  & 0.0339  & 0.0366  & 0.0472  & 0.0689  & 0.0445  & 0.0545  & 0.0566  & 0.0747  & 0.0578  & 0.0646  & 0.0770  & 0.1412  & 0.0745  & 0.0981 \\
UniCDR & 0.0359  & 0.0485  & 0.0380  & 0.0424  & 0.0493  & 0.0700  & 0.0502  & 0.0581  & 0.0589  & 0.0851  & 0.0607  & 0.0697  & 0.0876  & 0.1620  & 0.0823  & 0.1102 \\ \midrule
\rowcolor{Gray} \cellcolor{white} DRIP & \textbf{0.0403}*  & \textbf{0.0535}*  & \textbf{0.0423}*  & \textbf{0.0472}*  & \textbf{0.0526}*  & \textbf{0.0758}*  & \textbf{0.0517}  & \textbf{0.0619}*  & \textbf{0.0677}*  & \textbf{0.0958}*  & \textbf{0.0699}*  & \textbf{0.0813}*  & \textbf{0.0936}*  & \textbf{0.1785}*  & \textbf{0.0873}*  & \textbf{0.1193}* \\ \midrule
\textit{Improv.} & 12.42\%  & 10.42\%  & 11.56\%  & 11.39\%  & 6.84\%  & 8.30\%  & 3.13\%  & 6.54\%  & 14.86\%  & 12.58\%  & 14.56\%  & 16.66\%  & 6.78\%  & 10.23\%  & 4.44\%  & 5.50\% \\ \bottomrule
\multirow{2}{*}{\textbf{Methods (P2)}}   & \multicolumn{4}{c}{\textbf{Home}}  & \multicolumn{4}{c}{\textbf{Grocery}} & \multicolumn{4}{c}{\textbf{Tools}} & \multicolumn{4}{c}{\textbf{Health}} \\  \cmidrule(l){2-5} \cmidrule(l){6-9} \cmidrule(l){10-13} \cmidrule(l){14-17}
  & \textbf{R@20} & \textbf{R@50} & \textbf{N@20} & \textbf{N@50} & \textbf{R@20}  & \textbf{R@50}  & \textbf{N@20} & \textbf{N@50} & \textbf{R@20} & \textbf{R@50} & \textbf{N@20} & \textbf{N@50} & \textbf{R@20}  & \textbf{R@50} & \textbf{N@20} & \textbf{N@50} \\ \midrule
BPRMF & 0.0363  & 0.0600  & 0.0328  & 0.0423  & 0.0781  & 0.1513  & 0.0655  & 0.0943  & 0.0981  & 0.1613  & 0.0647  & 0.0873  & 0.0466  & 0.0735  & 0.0415  & 0.0525\\ \midrule
MMOE & 0.0305  & 0.0512  & 0.0279  & 0.0355  & 0.0966  & 0.1833  & 0.0928  & 0.1250  & 0.0799  & 0.1587  & 0.0678  & 0.0929  & 0.0494  & 0.0730  & 0.0520  & 0.0610 \\
PLE & 0.0279  & 0.0487  & 0.0279  & 0.0354  & 0.0953  & 0.1850  & 0.0912  & 0.1251  & 0.0814  & 0.1510  & 0.0697  & 0.0918  & 0.0500  & 0.0765  & 0.0525  & 0.0623 \\ \midrule
EMCDR+ & 0.0464  & 0.0795  & 0.0423  & 0.0552  & 0.1049  & 0.1982  & 0.0968  & 0.1336  & 0.1045  & 0.1727  & 0.0827  & 0.1097  & 0.0640  & 0.1032  & 0.0639  & 0.0791 \\ 
PTUPCDR+ & 0.0460  & 0.0788  & 0.0433  & 0.0557  & 0.1040  & 0.1945  & 0.0969  & 0.1329  & 0.1108  & 0.1857  & 0.0836  & 0.1100  & 0.0653  & 0.1050  & 0.0651  & 0.0810 \\ \midrule
CAT-ART+ & 0.0434  & 0.0801  & 0.0396  & 0.0535  & 0.1115  & 0.1938  & 0.0975  & 0.1286  & 0.1031  & 0.1758  & 0.0758  & 0.1018  & 0.0603  & 0.0967  & 0.0573  & 0.0712 \\
UniCDR &0.0468  & 0.0812  & 0.0436  & 0.0583  & 0.1181  & 0.2060  & 0.1068  & 0.1356  & 0.1206  & 0.1893  & 0.0863  & 0.1139  & 0.0674  & 0.1074  & 0.0647  & 0.0805 \\ \midrule
\rowcolor{Gray} \cellcolor{white} DRIP & \textbf{0.0527}*  & \textbf{0.0857}*  & \textbf{0.0472}*  & \textbf{0.0598}  & \textbf{0.1184}  & \textbf{0.2195}*  & \textbf{0.1086}*  & \textbf{0.1471}*  & \textbf{0.1229}*  & \textbf{0.1879}*  & \textbf{0.0919}*  & \textbf{0.1156}*  & \textbf{0.0726}*  & \textbf{0.1129}*  & \textbf{0.0715}*  & \textbf{0.0870}*\\ \midrule
\textit{Improv.} & 12.57\%  & 5.53\% & 8.33\%  & 2.63\%  & 2.70\%  & 6.52\%  & 3.23\%  & 8.48\%  & 11.10\%  & 6.08\%  & 14.97\%  & 10.75\%  & 7.78\%  & 5.07\%  & 9.82\%  & 7.40\% \\ \bottomrule 
\end{tabular}%
}
\scriptsize{* denotes the improvement over the best baseline is statistically significant with $p<0.05$, using the paired t-test.}
\caption{Recommendation performance comparison for MDRAU-ST on the platform scenario 1 (P1) and scenario 2 (P2). }
\label{table:many_to_one1}
\end{table*}

\subsection{Performance comparison for MDRAU-ST}
\label{subsec:exp_many_to_one}
We first compare the recommendation performance of various methods for MDRAU-ST on our two simulated platforms, P1 and P2.
In this task, we set each of the domains as the target domain, and for evaluation, we consider only the users who have not tried the domain yet as test users.
The results for each target domain are reported in Table~\ref{table:many_to_one1}. 

We observe that \proposed consistently achieves higher recommendation performance than all the other methods in each of the target domains.
Specifically, the CDR methods designed for a single pair of source-target domain based on one-to-one mapping (i.e., EMCDR+ and PTUPCDR+) show lower performance compared to the ones that can effectively handle multiple source domains based on many-to-one mapping (i.e., CAT-ART+, UniCDR, and \proposed).
They are not capable of capturing the relevance among multiple source domains, whereas the multi-domain CDR methods integrate the user preferences of multiple source domains in advanced ways.
In particular, the multi-domain encoder of \proposed contextualizes domain-specific user embeddings over multiple source domains with attention mechanism that is effective in capturing the inter-domain relationship;
this brings a significant performance improvement for MDRAU-ST.

Furthermore, unlike the CDR methods for recommending items in a user's unseen domain (i.e., EMCDR+, PTUPCDR+, and \proposed),
the CDR baselines that aim to enhance the recommendation in the user's seen domains (i.e., CAT-ART+ and UniCDR) have to rely on each user's global embedding shared over multiple domains to predict the user preference for a target unseen domain.
This is because they are suffering from the incapability of inferring a user's unseen domain embedding, caused by their training process aiming at only predicting items in a user's seen domains. 
In contrast, \proposed is good at inferring user embedding of a target unseen domain, conferred by masked domain modeling.
To sum up, \proposed outperforms all existing methods in terms of MDRAU-ST, by (1) effectively contextualizing user embeddings over multiple domains using attention mechanism and (2) accurately inferring a user's preference for a target unseen domain via masked domain modeling.

\subsection{Performance Comparison for \scenario-MT}
\label{subsec:main_table}
We also evaluate \proposed and various methods for \scenario-MT on two simulated platforms, P1 and P2.
In this task, the set of target domains varies depending on users, because each user has different unseen domains.
For each user, we assess the accuracy of recommendations for the user's multiple unseen domains, and report the averaged accuracy over all test users as the final performance.
Note that MDRAU-MT scenario differs from MDRAU-ST (in Sec.~\ref{subsec:exp_many_to_one}) in that it evaluates the models' ability to make aggregated recommendations over multiple unseen domains.
In this sense, \scenario-MT requires to properly capture \textit{domain-level preference} with item-level preference.
Table~\ref{table:main} presents overall performances for MDRAU-MT.

We observe that \proposed outperforms the best baseline method more in \scenario-MT than in \scenario-ST.
The limited MDRAU-MT performance of baselines stems from two factors.
First, they focus on predicting items in single domains without considering multi-domain preferences, unlike \proposed, which captures both domain and item-level preferences. 
Second, most baseline methods need a post-processing step to merge recommended item lists over multiple target domains, limiting final accuracy. 
On the contrary, \proposed optimizes a unified model that makes recommendations for multiple target domains in an end-to-end manner, improving MDRAU-MT performance by reducing its gap from the training process.
In conclusion, \proposed achieves the best performance among all the baselines with the help of its item-level preference accurately predicted for each of the target domains as well as its capability of inferring domain-level preference obtained by the training process.

\subsection{Domain-level Preference Analysis}
\label{subsec:domain-level}
We analyze the models' ability to capture domain-level preference, essential for accurate recommendations in multiple unseen domains (i.e., MDRAU-MT).
We compare two domain distributions obtained from (1) a user's interaction history and (2) the recommendation list generated by each method for the user.\footnote{In this analysis, we assume that the domain distributions in each user's interaction history reflect the user's actual domain-level preferences to a considerable extent.}
Let $P$ and $Q$ denote the ground-truth distribution in user history and the predicted distribution, respectively.
We calculate the Kullback-Leibler Divergence (KLD@$K$) between the two distributions $D_{KLD}(P\Vert Q)=\sum_{i} P_i\log (P_i / Q_i)$ to measure how closely the model's prediction captures the distribution of actual domain preferences.
Note that KLD only measures the domain-level accuracy, not in-domain item prediction accuracy.

\begin{figure}[t]
    \centering
    \includegraphics[width=0.99\columnwidth]{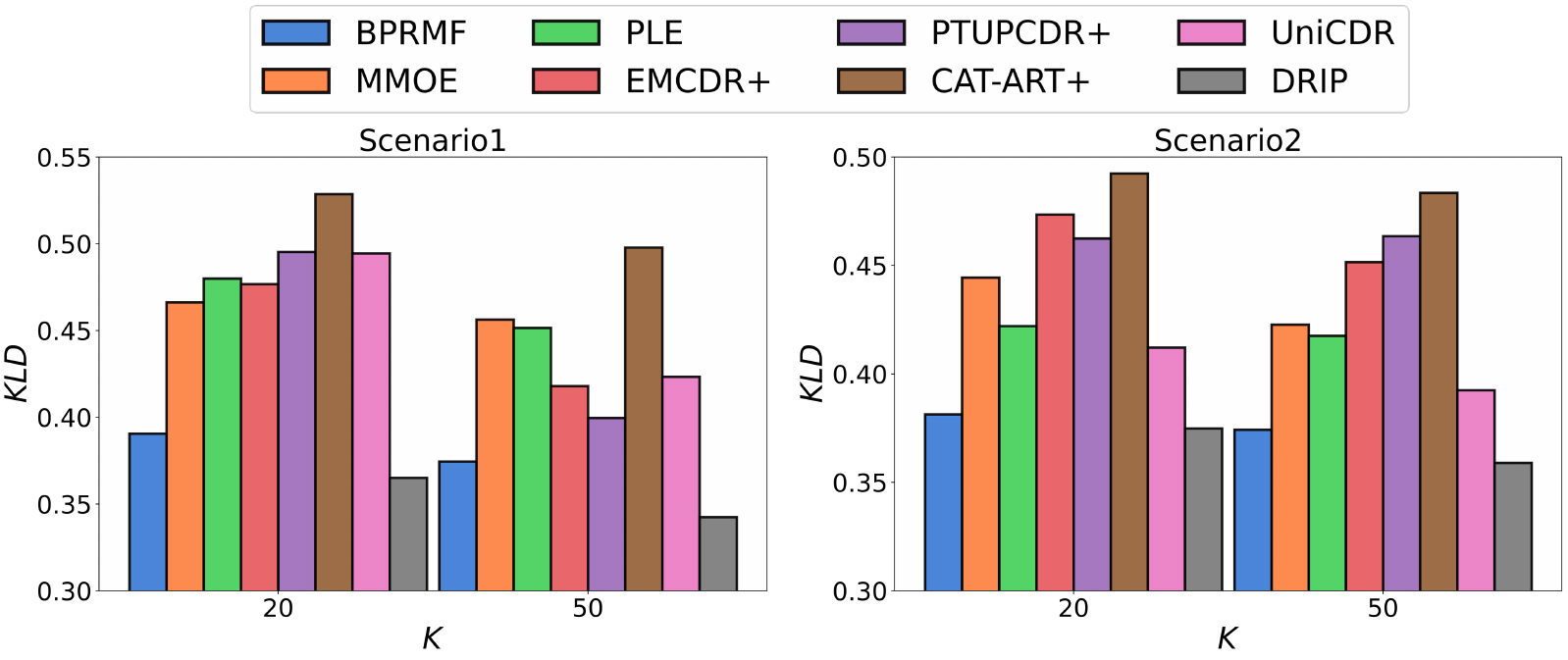}
    \caption{Domain-level preference analysis. KL-divergence scores of each method (Best viewed in color).}
    \label{fig:kld}
\end{figure}

In Fig.~\ref{fig:kld}, we assess the KLD@$K$ score for the top-$K$ recommendation and observe the following:
(1) In comparison to other competing methods, BPRMF exhibits significantly better KLD scores.
BPRMF treats a union of all domains as a single domain and learns the pair-wise ranking of item-level preference over the entire domains.
This makes a user's domain-level preference implicitly captured during the training process.
This result can be also interpreted with the previous performance comparison for MDRAU-MT/ST.
The performance of BPRMF is highly limited in MDRAU-ST due to its limited capability of capturing item-level preferences.
However, BPRMF achieves comparable performance with the state-of-the-art methods in MDRAU-MT.
We interpret that this improvement mainly comes from its capability of capturing domain-level preferences.
(2) Other competing methods show considerably worse KLD scores compared to \proposed.
As discussed earlier, they need a post-processing step to generate the unified ranking list encompassing all unseen domains (e.g., z-score normalization), which may not yield optimal recommendation accuracy.
This result highlights the importance of \textit{holistic} model training that considers both domain-level and item-level preferences in MDRAU task.
(3) Among all baseline methods, \proposed achieves the best KLD scores in both scenarios.
This result shows that \proposed indeed effectively captures user's domain-level preferences, and also supports its superior performance in MDRAU-MT.
\begin{table}[]
\resizebox{\columnwidth}{!}{%
\begin{tabular}{@{}ccccccccc@{}}
\toprule
\multirow{2}{*}{\textbf{Methods}} & \multicolumn{4}{c}{\textbf{P1}}    & \multicolumn{4}{c}{\textbf{P2}}    \\ \cmidrule(l){2-5} \cmidrule(l){6-9} 
                         & \textbf{R@20} & \textbf{R@50} & \textbf{N@20} & \textbf{N@50} & \textbf{R@20} & \textbf{R@50} & \textbf{N@20} & \textbf{N@50} \\  \midrule 
BPRMF                    & 0.0249  & 0.0315  & 0.0256  & 0.0278  & 0.0528  & 0.0835  & 0.0503  & 0.0625      \\ \midrule
MMOE                         & 0.0188  & 0.0230  & 0.0191  & 0.0203  & 0.0486  & 0.0691  & 0.0513  & 0.0581    \\
PLE                         & 0.0190  & 0.0249  & 0.0190  & 0.0212  & 0.0491  & 0.0722  & 0.0516  & 0.0595     \\ \midrule
EMCDR+                         & 0.0362  & 0.0507  & 0.0374  & 0.0434  & 0.0531  & 0.0832  & 0.0525  & 0.0659     \\
PTUPCDR+                       & 0.0367  & 0.0482  & 0.0377  & 0.0422  & 0.0546  & 0.0866  & 0.0538  & 0.0665    \\ \midrule
CAT-ART+                         &0.0395  & 0.0517  & 0.0395  & 0.0441  & 0.0522  & 0.0827  & 0.0497  & 0.0620   \\
UniCDR                         & 0.0434  & 0.0571  & 0.0446  & 0.0497  & 0.0588  & 0.0958  & 0.0574  & 0.0719   \\ \midrule
\rowcolor{Gray} \cellcolor{white}DRIP                         &   \textbf{0.0545}*  & \textbf{0.0698}*  & \textbf{0.0556}*  & \textbf{0.0612}*  & \textbf{0.0780}*  & \textbf{0.1170}*  & \textbf{0.0773}*  & \textbf{0.0919}*     \\ \midrule
\textit{Improv.}                         & 25.64\%  & 22.25\%  & 24.55\%  & 23.14\%  & 32.66\%  & 22.11\%  & 34.80\%  & 27.78\%    \\ \bottomrule
\end{tabular}%
}
\scriptsize{* denotes the improvement over the best baseline is statistically significant with $p<0.05$, using the paired t-test.}
\caption{Recommendation performance comparison for \scenario-MT.}
\label{table:main}
\end{table}

\begin{table}[t]
\resizebox{\columnwidth}{!}{%
\begin{tabular}{@{}lcccc@{}}
\toprule
\textbf{Designs}       & \textbf{R@20} & \textbf{R@50} & \textbf{N@20} & \textbf{N@50} \\ \midrule\midrule
\textbf{\proposed} & \textbf{0.0545}      &  \textbf{0.0698}    & \textbf{0.0556}     & \textbf{0.0612} \\ \midrule\midrule
\;\:\,\textbf{Training Paradigm} \\ \midrule
\quad Single-domain learning        &     0.0281  & 0.0371  & 0.0272  & 0.0310 \\ 
\quad Many-to-one learning w/ post-processing A  & 0.0432  & 0.0593  & 0.0447  & 0.0511\\
\quad Many-to-one learning w/ post-processing B & 0.0409  & 0.0520  & 0.0424  & 0.0463\\
\midrule\midrule
\multicolumn{5}{l}{\textbf{Domain Preference Modeling}} \\ \midrule
\quad Uniform dist.       & 0.0122  & 0.0210  & 0.0120  & 0.0164         \\
\quad Domain activeness dist.         &    0.0409  & 0.0534  & 0.0419  & 0.0466     \\ \midrule\midrule
\multicolumn{5}{l}{\textbf{Masking Scheme}} \\ \midrule
\quad w/o Adaptive Masking        &  0.0513     &0.0663      & 0.0520     & 0.0576             \\  \bottomrule
\end{tabular}%
}
\caption{Performance comparison of different design choices. Results for MDRAU-MT on P1.}
\label{table:ablation}
\end{table}

\subsection{Design Choice Analysis}
We analyze alternative design choices for \proposed in Table~\ref{table:ablation} to verify the effectiveness of our design choice.
We report the performance for MDRAU-MT on platform scenario P1.

First, we compare alternative training paradigms for \proposed:
Single-domain learning, where all domains are treated as one, and the model is trained to maximize the likelihood of the training data; Many-to-one learning, where a single model is trained for each target domain to predict item-level preferences.
To generate a unified recommendation list encompassing all domains, we apply a post-processing step.
Post-processing A uses z-normalization, performing best in our test.
We also consider emphasizing active domains in post-processing; specifically, for post-processing B, we multiply the ratio of the total number of interactions in each domain by the z-normalized scores and use the scores for the recommendation.

We observe that the alternative training paradigms show considerably degraded performance.
Single-domain learning achieves a highly limited performance, showing the necessity of proper domain modeling in MDRAU task.
Also, the many-to-one learning neglects the domain-level preference during the training, and the post-processing is applied independently from the training process, which results in limited MDRAU performance.
Further, in our experiments, sophisticated designs for post-processing do not bring further improvements.
These results support the superiority of our training strategy that decomposes user preference into domain- and item-level preferences and jointly learns them through a unified model.

Second, we compare ablations for the domain-level preference modeling.
Instead of estimating personalized domain-level preference, they use globally fixed distributions:
the uniform and domain activeness distribution (the latter assumes users prefer more active domains).
Both fixed domain preference approaches yield suboptimal recommendation performance.
This result supports the effectiveness of our strategy that models the domain-level preference for each individual user considering their different preference.

Lastly, we provide the results without the adaptive masking.
For the ablation, we use random masking with the same masking ratio as adaptive masking.
The adaptive masking brings slight improvements to the final performance, indicating that our masking strategy is well-aligned with our masked domain modeling.
We also provide the sensitivity analysis in the Appendix.
\section{Conclusion}
This paper highlights the importance of \scenario task based on its practical advantages in multi-domain platforms.
We propose \proposed, a new framework to provide accurate unseen domain recommendations to attract users into new service domains that they have not interacted with yet.
The \proposed decomposes user preference into domain-level preference and in-domain item-level preference and then jointly learns them via a unified model with the help of a training strategy based on masked domain modeling.
We conduct extensive comparisons with a wide range of CDR methods.
\proposed consistently achieves superior performance compared to all competing methods in both cases of a specific target domain (\scenario-ST) and multiple target domains (\scenario-MT).
We expect that \proposed can enhance the user experience by fostering diverse and serendipitous discoveries and potentially promote the influx of new users to each service domain, benefiting providers in multi-domain service platforms.

\section*{Acknowledgements}
This work was supported by Institute of Information \& communications Technology Planning \& Evaluation (IITP) grant funded by the Korea government (MSIT) (No.2018-0-00584, No.2019-0-01906), and the National Research Foundation of Korea (NRF) grant funded by the MSIT (No.2020R1A2B5B03097210, No.RS-2023-00217286).
\bibliography{ref}

\begin{thebibliography}{39}
\providecommand{\natexlab}[1]{#1}

\bibitem[{Ba, Kiros, and Hinton(2016)}]{layernorm}
Ba, J.~L.; Kiros, J.~R.; and Hinton, G.~E. 2016.
\newblock Layer normalization.
\newblock \emph{arXiv preprint arXiv:1607.06450}.

\bibitem[{Bao et~al.(2022)Bao, Dong, Piao, and Wei}]{beit}
Bao, H.; Dong, L.; Piao, S.; and Wei, F. 2022.
\newblock {BE}iT: {BERT} Pre-Training of Image Transformers.
\newblock In \emph{International Conference on Learning Representations}.

\bibitem[{Cao et~al.(2023)Cao, Li, Yu, Guo, Liu, and Wang}]{unicdr}
Cao, J.; Li, S.; Yu, B.; Guo, X.; Liu, T.; and Wang, B. 2023.
\newblock Towards Universal Cross-Domain Recommendation.
\newblock In \emph{Proceedings of the Sixteenth ACM International Conference on Web Search and Data Mining}, 78–86.

\bibitem[{Cao et~al.(2022{\natexlab{a}})Cao, Lin, Cong, Ya, Liu, and Wang}]{disencdr}
Cao, J.; Lin, X.; Cong, X.; Ya, J.; Liu, T.; and Wang, B. 2022{\natexlab{a}}.
\newblock DisenCDR: Learning Disentangled Representations for Cross-Domain Recommendation.
\newblock In \emph{Proceedings of the 45th International ACM SIGIR Conference on Research and Development in Information Retrieval}, 267--277.

\bibitem[{Cao et~al.(2022{\natexlab{b}})Cao, Sheng, Cong, Liu, and Wang}]{cdrib}
Cao, J.; Sheng, J.; Cong, X.; Liu, T.; and Wang, B. 2022{\natexlab{b}}.
\newblock Cross-Domain Recommendation to Cold-Start Users via Variational Information Bottleneck.
\newblock In \emph{2022 IEEE 38th International Conference on Data Engineering (ICDE)}, 2209--2223.

\bibitem[{Cui et~al.(2020)Cui, Wei, Zhang, and Zhang}]{cui2020herograph}
Cui, Q.; Wei, T.; Zhang, Y.; and Zhang, Q. 2020.
\newblock HeroGRAPH: A Heterogeneous Graph Framework for Multi-Target Cross-Domain Recommendation.
\newblock In \emph{ORSUM@ RecSys}.

\bibitem[{Devlin et~al.(2019)Devlin, Chang, Lee, and Toutanova}]{BERT}
Devlin, J.; Chang, M.-W.; Lee, K.; and Toutanova, K. 2019.
\newblock {BERT}: Pre-training of Deep Bidirectional Transformers for Language Understanding.
\newblock In \emph{Proceedings of the 2019 Conference of the North {A}merican Chapter of the Association for Computational Linguistics: Human Language Technologies, Volume 1 (Long and Short Papers)}, 4171--4186.

\bibitem[{Fu et~al.(2019)Fu, Peng, Wang, Xu, and Li}]{deeplyfusing}
Fu, W.; Peng, Z.; Wang, S.; Xu, Y.; and Li, J. 2019.
\newblock Deeply fusing reviews and contents for cold start users in cross-domain recommendation systems.
\newblock In \emph{Proceedings of the AAAI Conference on Artificial Intelligence}, 94--101.

\bibitem[{He and McAuley(2016)}]{amazon}
He, R.; and McAuley, J. 2016.
\newblock Ups and Downs: Modeling the Visual Evolution of Fashion Trends with One-Class Collaborative Filtering.
\newblock In \emph{Proceedings of the 25th International Conference on World Wide Web}, WWW '16, 507–517. Republic and Canton of Geneva, CHE: International World Wide Web Conferences Steering Committee.
\newblock ISBN 9781450341431.

\bibitem[{He et~al.(2020)He, Deng, Wang, Li, Zhang, and Wang}]{lgcn}
He, X.; Deng, K.; Wang, X.; Li, Y.; Zhang, Y.; and Wang, M. 2020.
\newblock Lightgcn: Simplifying and powering graph convolution network for recommendation.
\newblock In \emph{Proceedings of the 43rd International ACM SIGIR conference on research and development in Information Retrieval}, 639--648.

\bibitem[{He et~al.(2017)He, Liao, Zhang, Nie, Hu, and Chua}]{he2017neural}
He, X.; Liao, L.; Zhang, H.; Nie, L.; Hu, X.; and Chua, T.-S. 2017.
\newblock Neural collaborative filtering.
\newblock In \emph{Proceedings of the 26th international conference on world wide web}, 173--182.

\bibitem[{Hsieh et~al.(2017)Hsieh, Yang, Cui, Lin, Belongie, and Estrin}]{cml}
Hsieh, C.-K.; Yang, L.; Cui, Y.; Lin, T.-Y.; Belongie, S.; and Estrin, D. 2017.
\newblock Collaborative Metric Learning.
\newblock 193–201. International World Wide Web Conferences Steering Committee.

\bibitem[{Hu, Zhang, and Yang(2018)}]{conet}
Hu, G.; Zhang, Y.; and Yang, Q. 2018.
\newblock CoNet: Collaborative Cross Networks for Cross-Domain Recommendation.
\newblock In \emph{Proceedings of the 27th ACM International Conference on Information and Knowledge Management}, 667–676.

\bibitem[{Kang et~al.(2019)Kang, Hwang, Lee, and Yu}]{sscdr}
Kang, S.; Hwang, J.; Lee, D.; and Yu, H. 2019.
\newblock Semi-supervised learning for cross-domain recommendation to cold-start users.
\newblock In \emph{Proceedings of the 28th ACM International Conference on Information and Knowledge Management}, 1563--1572.

\bibitem[{Kingma and Ba(2014)}]{adam}
Kingma, D.~P.; and Ba, J. 2014.
\newblock Adam: A method for stochastic optimization.
\newblock \emph{arXiv preprint arXiv:1412.6980}.

\bibitem[{Koren, Bell, and Volinsky(2009)}]{koren_mf}
Koren, Y.; Bell, R.; and Volinsky, C. 2009.
\newblock Matrix Factorization Techniques for Recommender Systems.
\newblock \emph{Computer}, 42(8): 30--37.

\bibitem[{Krichene and Rendle(2020)}]{fulleval}
Krichene, W.; and Rendle, S. 2020.
\newblock On Sampled Metrics for Item Recommendation.
\newblock KDD '20, 1748–1757. New York, NY, USA: Association for Computing Machinery.
\newblock ISBN 9781450379984.

\bibitem[{Li et~al.(2023)Li, Xie, Yu, Hu, Li, Shu, Qie, and Niu}]{cat_art}
Li, C.; Xie, Y.; Yu, C.; Hu, B.; Li, Z.; Shu, G.; Qie, X.; and Niu, D. 2023.
\newblock One for All, All for One: Learning and Transferring User Embeddings for Cross-Domain Recommendation.
\newblock In \emph{Proceedings of the Sixteenth ACM International Conference on Web Search and Data Mining}, 366–374.

\bibitem[{Li and Tuzhilin(2020)}]{ddtcdr}
Li, P.; and Tuzhilin, A. 2020.
\newblock Ddtcdr: Deep dual transfer cross domain recommendation.
\newblock In \emph{Proceedings of the 13th International Conference on Web Search and Data Mining}, 331--339.

\bibitem[{Liang et~al.(2018)Liang, Krishnan, Hoffman, and Jebara}]{vaecf}
Liang, D.; Krishnan, R.~G.; Hoffman, M.~D.; and Jebara, T. 2018.
\newblock Variational autoencoders for collaborative filtering.
\newblock In \emph{Proceedings of the 2018 world wide web conference}, 689--698.

\bibitem[{Liu et~al.(2020)Liu, Li, Li, and Pan}]{bitgcf}
Liu, M.; Li, J.; Li, G.; and Pan, P. 2020.
\newblock Cross domain recommendation via bi-directional transfer graph collaborative filtering networks.
\newblock In \emph{Proceedings of the 29th ACM International Conference on Information \& Knowledge Management}, 885--894.

\bibitem[{Ma et~al.(2018)Ma, Zhao, Yi, Chen, Hong, and Chi}]{mmoe}
Ma, J.; Zhao, Z.; Yi, X.; Chen, J.; Hong, L.; and Chi, E.~H. 2018.
\newblock Modeling task relationships in multi-task learning with multi-gate mixture-of-experts.
\newblock In \emph{Proceedings of the 24th ACM SIGKDD international conference on knowledge discovery \& data mining}, 1930--1939.

\bibitem[{Man et~al.(2017)Man, Shen, Jin, and Cheng}]{emcdr}
Man, T.; Shen, H.; Jin, X.; and Cheng, X. 2017.
\newblock Cross-Domain Recommendation: An Embedding and Mapping Approach.
\newblock In \emph{Proceedings of the Twenty-Sixth International Joint Conference on Artificial Intelligence, {IJCAI-17}}, 2464--2470.

\bibitem[{Paszke et~al.(2019)Paszke, Gross, Massa, Lerer, Bradbury, Chanan, Killeen, Lin, Gimelshein, Antiga, Desmaison, Kopf, Yang, DeVito, Raison, Tejani, Chilamkurthy, Steiner, Fang, Bai, and Chintala}]{pytorch}
Paszke, A.; Gross, S.; Massa, F.; Lerer, A.; Bradbury, J.; Chanan, G.; Killeen, T.; Lin, Z.; Gimelshein, N.; Antiga, L.; Desmaison, A.; Kopf, A.; Yang, E.; DeVito, Z.; Raison, M.; Tejani, A.; Chilamkurthy, S.; Steiner, B.; Fang, L.; Bai, J.; and Chintala, S. 2019.
\newblock PyTorch: An Imperative Style, High-Performance Deep Learning Library.
\newblock In Wallach, H.; Larochelle, H.; Beygelzimer, A.; d\textquotesingle Alch\'{e}-Buc, F.; Fox, E.; and Garnett, R., eds., \emph{Advances in Neural Information Processing Systems}, volume~32. Curran Associates, Inc.

\bibitem[{Rendle et~al.(2009)Rendle, Freudenthaler, Gantner, and Schmidt-Thieme}]{bpr}
Rendle, S.; Freudenthaler, C.; Gantner, Z.; and Schmidt-Thieme, L. 2009.
\newblock BPR: Bayesian Personalized Ranking from Implicit Feedback.
\newblock In \emph{Proceedings of the Twenty-Fifth Conference on Uncertainty in Artificial Intelligence}, 452–461.

\bibitem[{Singh and Gordon(2008)}]{kdd08}
Singh, A.~P.; and Gordon, G.~J. 2008.
\newblock Relational learning via collective matrix factorization.
\newblock In \emph{Proceedings of the 14th ACM SIGKDD international conference on Knowledge discovery and data mining}, 650--658.

\bibitem[{Srivastava et~al.(2014)Srivastava, Hinton, Krizhevsky, Sutskever, and Salakhutdinov}]{dropout}
Srivastava, N.; Hinton, G.; Krizhevsky, A.; Sutskever, I.; and Salakhutdinov, R. 2014.
\newblock Dropout: a simple way to prevent neural networks from overfitting.
\newblock \emph{The journal of machine learning research}, 15(1): 1929--1958.

\bibitem[{Tang et~al.(2020)Tang, Liu, Zhao, and Gong}]{ple}
Tang, H.; Liu, J.; Zhao, M.; and Gong, X. 2020.
\newblock Progressive layered extraction (ple): A novel multi-task learning (mtl) model for personalized recommendations.
\newblock In \emph{Proceedings of the 14th ACM Conference on Recommender Systems}, 269--278.

\bibitem[{Vaswani et~al.(2017)Vaswani, Shazeer, Parmar, Uszkoreit, Jones, Gomez, Kaiser, and Polosukhin}]{transformer}
Vaswani, A.; Shazeer, N.; Parmar, N.; Uszkoreit, J.; Jones, L.; Gomez, A.~N.; Kaiser, L.~u.; and Polosukhin, I. 2017.
\newblock Attention is All you Need.
\newblock In \emph{Advances in Neural Information Processing Systems 2017}, volume~30.

\bibitem[{Wang et~al.(2019)Wang, He, Wang, Feng, and Chua}]{ngcf}
Wang, X.; He, X.; Wang, M.; Feng, F.; and Chua, T.-S. 2019.
\newblock Neural Graph Collaborative Filtering.
\newblock In \emph{Proceedings of the 42nd International ACM SIGIR Conference on Research and Development in Information Retrieval}, 165–174.

\bibitem[{Yuan, Yao, and Benatallah(2019)}]{darec}
Yuan, F.; Yao, L.; and Benatallah, B. 2019.
\newblock DARec: Deep domain adaptation for cross-domain recommendation via transferring rating patterns.
\newblock \emph{arXiv preprint arXiv:1905.10760}.

\bibitem[{Zhang, Cao, and Yeung(2010)}]{mcf}
Zhang, Y.; Cao, B.; and Yeung, D.-Y. 2010.
\newblock Multi-Domain Collaborative Filtering.
\newblock In \emph{Proceedings of the Twenty-Sixth Conference on Uncertainty in Artificial Intelligence}, 725–732.

\bibitem[{Zhao, Yang, and Yu(2022)}]{msdcr}
Zhao, X.; Yang, N.; and Yu, P.~S. 2022.
\newblock Multi-Sparse-Domain Collaborative Recommendation via Enhanced Comprehensive Aspect Preference Learning.
\newblock In \emph{Proceedings of the Fifteenth ACM International Conference on Web Search and Data Mining}, 1452--1460.

\bibitem[{Zhu et~al.(2019)Zhu, Chen, Wang, Liu, and Zheng}]{dtcdr}
Zhu, F.; Chen, C.; Wang, Y.; Liu, G.; and Zheng, X. 2019.
\newblock Dtcdr: A framework for dual-target cross-domain recommendation.
\newblock In \emph{Proceedings of the 28th ACM International Conference on Information and Knowledge Management}, 1533--1542.

\bibitem[{Zhu et~al.(2020)Zhu, Wang, Chen, Liu, and Zheng}]{GA-dtcdr}
Zhu, F.; Wang, Y.; Chen, C.; Liu, G.; and Zheng, X. 2020.
\newblock A Graphical and Attentional Framework for Dual-Target Cross-Domain Recommendation.
\newblock In \emph{Proceedings of the Twenty-Ninth International Joint Conference on Artificial Intelligence, {IJCAI-20}}, 3001--3008.

\bibitem[{Zhu et~al.(2021{\natexlab{a}})Zhu, Wang, Chen, Zhou, Li, and Liu}]{cdrsurvey}
Zhu, F.; Wang, Y.; Chen, C.; Zhou, J.; Li, L.; and Liu, G. 2021{\natexlab{a}}.
\newblock Cross-Domain Recommendation: Challenges, Progress, and Prospects.
\newblock In \emph{Proceedings of the Thirtieth International Joint Conference on Artificial Intelligence, {IJCAI-21}}, 4721--4728.

\bibitem[{Zhu et~al.(2023)Zhu, Wang, Zhou, Chen, Li, and Liu}]{ga_mtcdr}
Zhu, F.; Wang, Y.; Zhou, J.; Chen, C.; Li, L.; and Liu, G. 2023.
\newblock A Unified Framework for Cross-Domain and Cross-System Recommendations.
\newblock \emph{IEEE Transactions on Knowledge and Data Engineering}, 35(2): 1171--1184.

\bibitem[{Zhu et~al.(2021{\natexlab{b}})Zhu, Ge, Zhuang, Xie, Xi, Zhang, Lin, and He}]{tmcdr}
Zhu, Y.; Ge, K.; Zhuang, F.; Xie, R.; Xi, D.; Zhang, X.; Lin, L.; and He, Q. 2021{\natexlab{b}}.
\newblock Transfer-Meta Framework for Cross-Domain Recommendation to Cold-Start Users.
\newblock In \emph{Proceedings of the 44th International ACM SIGIR Conference on Research and Development in Information Retrieval}, 1813–1817.

\bibitem[{Zhu et~al.(2022)Zhu, Tang, Liu, Zhuang, Xie, Zhang, Lin, and He}]{ptupcdr}
Zhu, Y.; Tang, Z.; Liu, Y.; Zhuang, F.; Xie, R.; Zhang, X.; Lin, L.; and He, Q. 2022.
\newblock Personalized transfer of user preferences for cross-domain recommendation.
\newblock In \emph{Proceedings of the Fifteenth ACM International Conference on Web Search and Data Mining}, 1507--1515.

\end{thebibliography}
\appendix
\clearpage

\section{Transformer Layer}
The multi-domain encoder is basically a stack of $L$ transformer layers \cite{transformer, BERT}, each of which consists of two blocks: multi-head self-attention (\msa) and position-wise two-layered feed-forward neural network (FFN).
Given the input representation $\bm{H}\in \mathbb{R}^{(K+1)\times m}$, the \msa block computes its output as follows:
\begin{equation}
\begin{split}
    \msa(\bm{H}) &= \left[\satth_1, \satth_2, \cdots, \satth_h\right]\bm{W}^O \\
    \satth_i &= \satt(\bm{H}\bm{W}_i^Q, \bm{H}\bm{W}_i^K, \bm{H}\bm{W}_i^V)\\
    \satt(\bm{Q}, \bm{K}, \bm{V}) &= \text{Softmax}\left(\bm{Q}\bm{K}^{\top}/\sqrt{m/h}\right)\bm{V},
\end{split}
\end{equation}
where $\bm{W}_i^Q, \bm{W}_i^K, \bm{W}_i^V\in \mathbb{R}^{m\times (m/h)}$ and $\bm{W}^O\in \mathbb{R}^{m\times m}$ are learnable parameters, and $h$ denotes the number of heads.
The output of each block is applied dropout~\cite{dropout}, then added to the block's input for the residual connection, and finally normalized with Layer Norm~\cite{layernorm}.
Transformer Layer calculates the \msa block, and then the output is forwarded to the FFN block.

\section{Algorithm}
For a better understanding of \proposed (Sec.~\ref{sec:proposed}), we provide the algorithms for \proposed.
Algorithm~\ref{alg:algorithm-train} shows the training procedure, while Algorithm~\ref{alg:algorithm-test} shows the procedure for making a recommendation list of a user for each scenario (i.e., MDRAU-MT/ST).

\begin{algorithm}[h]
\caption{DRIP Training}
\label{alg:algorithm-train}
\textbf{Input}: Interaction data $\bm{R}$, $K$ domain-specific encoders (i.e., RSs) (Sec.~\ref{subsec:encoder})\\
\textbf{Output}: A trained \proposed model
  \begin{algorithmic}[1]
  \WHILE{\textit{not converged}}
    \FOR{\textit{each batch users} $\mathcal{B}(u)\in\mathcal{U}$}
        \STATE Construct inputs according to users' interaction \\
        information (Eq.~\eqref{eq:input_emb})
        \STATE Calculate the domain-level preferences: \\
        $p(d_k;u;\bm{\Theta})$ (Eq.~\eqref{eq:dom-level-pref})        
        \STATE Select masking probability with selection probability (Sec.~\ref{subsec:learning},~\ref{subsec:impl-detail})
        \STATE Mask inputs for training (Eq.~\eqref{eq:masking-train})
        \STATE Compute $\mathcal{L}_{\proposed}$ (Eq.~\eqref{eq:loss})
        \STATE Update model by minimizing $\mathcal{L}_{\proposed}$
    \ENDFOR
  \ENDWHILE
  \end{algorithmic}
\end{algorithm}

\begin{algorithm}[h]
\caption{DRIP Inference}
\textbf{Input}: A trained \proposed model, a user $u$, and recommendation size $N$ \\
\textbf{Output}: A recommendation list
\label{alg:algorithm-test}
  \begin{algorithmic}[1]
    \STATE Construct an input according to the user's interaction information (Eq.~\eqref{eq:input_emb})
    \STATE Calculate $p(d_k|u;\bm{\Theta})$ and $p(v|u,d_k;\bm{\Theta})$ (Eq.~\eqref{eq:dom-level-pref},~\eqref{eq:item-level-preference})
    \IF{\scenario-MT Scenario}
        \STATE Calculate the preferences: $p(d_k|u;\bm{\Theta}) \cdot p(v|u,d_k;\bm{\Theta})$
        \STATE Sort the preferences and select top-$N$ items for top-$N$ recommendation
    \ELSIF{\scenario-ST Scenario (domain $d_k$)}
        \STATE Calculate the preferences: $p(v|u,d_k;\bm{\Theta})$
        \STATE Sort the preferences and select top-$N$ items for top-$N$ recommendation for domain $d_k$
    \ENDIF
  \end{algorithmic}
\end{algorithm}

\section{Detailed Descriptions for Experiments}
\subsection{Experimental Settings}
\subsubsection{Dataset and Domain Settings}
We filter out users who interacted with at least two domains with less than 10 (P1) \& 5 (P2) interactions in each domain and users who interacted with only one domain and items with less than 20 (P1) \& 10 (P2) interactions in each domain~\cite{sscdr, deeplyfusing}.
We report overlapping user information for our datasets in Table~\ref{table:overlap}.

\subsubsection{Evaluation Setup}
To assess MDRAU accuracy, we remove interaction history of overlapping users in certain domains and treat the removed interactions as ground-truth for evaluation.
In specific, for each overlapping user, we randomly select a subset of interacted domains and hide all interactions in the selected domains\footnote{We select the domain to be hidden with the probability of $0.3$. Note that the hidden interactions are completely excluded from the training data.}.
The selected domains are regarded as ground-truth unseen domains that the user is likely to interact with, and the hidden interacted items are regarded as ground-truth items in the domains.
For MDRAU-ST, we evaluate the quality of recommendation for each unseen domain, respectively, and for MDRAU-MT, we evaluate the quality of recommendation encompassing all unseen domains.

\subsubsection{Evaluation Metrics}
\begin{table}[tbp]
\centering
\resizebox{0.7\columnwidth}{!}{
\begin{tabular}{@{}cccccc@{}}
\toprule
                  & Domains    & Book  & Movie  & CD      & Game  \\ \midrule
\multirow{4}{*}{P1} & Book    & 35,987 & 5,243   & 1,678    & 316   \\
                            & Movie   & -     & 17,056  & 2,884    & 548   \\
                            & CD      & -     & -      & 5,941    & 222   \\
                            & Game    & -     & -      & -       & 1,049  \\\midrule
                   & Domains    & Home  & Health & Grocery & Tools \\ \midrule
\multirow{4}{*}{P2} & Home    & 10,317 & 5,444   & 3,072    & 1,399      \\
                            & Health  &  -     & 8,690   & 3,441    & 1,338    \\
                            & Grocery &  -    &   -     & 4,857    & 891      \\
                            & Tools   &  -     & -       &  -       & 1,804      \\ \bottomrule
\end{tabular}}
\caption{Count of users who interact with each domain pair.}
\label{table:overlap}
\end{table}
We focus on the top-$K$ recommendation task for implicit feedback.
We evaluate the recommendation accuracy of each method by using the two ranking metrics~\cite{cml,vaecf}: Recall (R@$K$) and Normalized Discounted Cumulative Gain (N@$K$).
R@$K$ equally treats all items within a top-$K$ list, while N@$K$ assigns a higher score on upper-ranked test items.
We use 50\% of overlapping users for validation purposes.
We use the full-ranking evaluation ~\cite{fulleval} rather than the sampling-based evaluation (e.g., leave-one-out~\cite{conet, sscdr, unicdr}).

\subsection{Compared Methods}
We compare \proposed with various methods from related research fields.
As discussed in the paper, MDRAU task has not been studied well in previous literature, and many existing methods cannot be directly applied. 
Therefore, we have modified the original methods to perform MDRAU, and the modified versions are annotated with the suffix `+'.
The following are our competing methods.

\begin{itemize}[leftmargin=*]
    \item \textbf{BPRMF~\cite{bpr}} is a representative method for implicit feedback.
    It learns pair-wise ranking in the latent space.
\end{itemize}

\begin{itemize}[leftmargin=*]
    \item \textbf{MMOE~\cite{mmoe}} utilizes multiple experts that share user embeddings over multiple domains, and it adopts a domain-specific gating module to effectively capture inter-domain relationships.
    \item \textbf{PLE~\cite{ple}} also utilizes multiple experts, but it separates domain-specific and domain-shared experts for more balanced training.
\end{itemize}
Note that the above methods learn global user embeddings that are shared across all domains, so that the global user embeddings can be directly used for recommendation in any target domain.

EMCDR and PTUPCDR learn a mapping function for each source-target domain pair.
Due to a large number of seen-unseen domain combinations, it is infeasible to directly apply them to MDRAU.
For this reason, we tailor their learning task for a many-to-one mapping.
That is, for each target domain, we learn a mapping function that takes \textit{concatenated} domain-specific user embeddings as input\footnote{If a user does not exist in a specific domain, we substitute the user embedding by the average user embedding in that domain.} and predicts user preference in the target domain.
\begin{itemize}[leftmargin=*]
    \item \textbf{EMCDR+~\cite{emcdr}} is a representative CDR method based on the embedding-and-mapping approach for unseen domain recommendation (or cold-start recommendation).
    It trains a mapping function to predict user embedding in the target domain from the source domain.
    The `+' version (i.e., EMCDR+) is modified according to the above description.

    \item \textbf{PTUPCDR+~\cite{ptupcdr}} improves EMCDR+ by employing a meta-network as its personalized mapping function.
    It trains the mapping function to predict ratings in the target domain.
\end{itemize}

\begin{itemize}[leftmargin=*]
    \item \textbf{CAT-ART+~\cite{cat_art}} generates shared user embeddings that summarize user preferences from all interacted domains and exploits the shared embeddings to improve the recommendation quality in the target domain.
    As user embeddings do not exist for unseen domains, we use the \textit{reconstructed user embedding} from the autoencoder module used to generate the shared embedding.
    
    \item \textbf{UniCDR~\cite{unicdr}} is the state-of-the-art CDR method that can handle various CDR scenarios.
    It models domain-specific and domain-shared user embeddings separately and transfers the knowledge from other domains based on the domain-shared embeddings.
\end{itemize}
All the compared methods except BPRMF were proposed to recommend items for a single target domain.
To evaluate such single target methods in the MDRAU-MT, the recommendations for each unseen domain need to be integrated over multiple unseen domains.
For effective integration, we generate a unified recommendation list by using the scores normalized within each domain.\footnote{We found that the performance of the single target methods drastically drops without normalization. We report their best performance among the cases of using z-score normalization and min-max normalization.}

\subsection{Implementation Details}
\label{subsec:impl-detail}
We use Xeon Gold 6226R CPU and A5000 24GB GPU on Ubuntu 20.04 LTS for experiments.
We implement \proposed and other competing methods with PyTorch~\cite{pytorch} and use Adam optimizer~\cite{adam} with $\beta_1=0.9, \beta_2=0.999$.
For domain-specific encoders, we use BPR \cite{bpr} with embedding size 64.
We ran five times for each experiment.
We tune all hyperparameters by grid search using the validation set.
The learning rates are searched in the range of $\{0.0001, 0.0005, 0.001, 0.005, 0.01\}$.
The weight decay is searched in the range of $\{0.0001, 0.0005, 0.001, 0.005, 0.01\}$.
For MMOE and PLE, the number of experts is searched in the range of $\{2, 3, 5, 8, 10\}$.
For the mapping function in EMCDR, we use multi-layer perception (MLP) with one-hidden layer: $[64 \times 4 \rightarrow 2\times 64\times 4 \rightarrow 64 \times 4]$ and $tanh$ activation \cite{emcdr, sscdr}.
For CAT-ART and UniCDR, we follow the recommended values from the public implementation and from the original papers~\cite{cat_art, unicdr}.
For \proposed, the number of heads, the number of layers, and random masking probabilities are searched in the range of $\{2^0, 2^1, 2^2, 2^3\}$, $\{1, 2, 3, 4\}$, $\{0,1, 0.2, 0.3, 0.4, 0.5, 0.6, 0.7, 0.9\}$, respectively. The results are reported in Sec.~\ref{subsec:sensitivity}.
For the masking process, we introduce a selection probability $\epsilon_i$ in epoch $i$.
When $\epsilon_i =1$, we use only random masking, while when $\epsilon_i=0$, we only use adaptive masking.
We linearly decrease the $\epsilon_i=\text{max}(0.5, 1- 0.002 i)$ in training phase.

\begin{figure}[t]
    \centering
    \includegraphics[width=1.0\columnwidth]{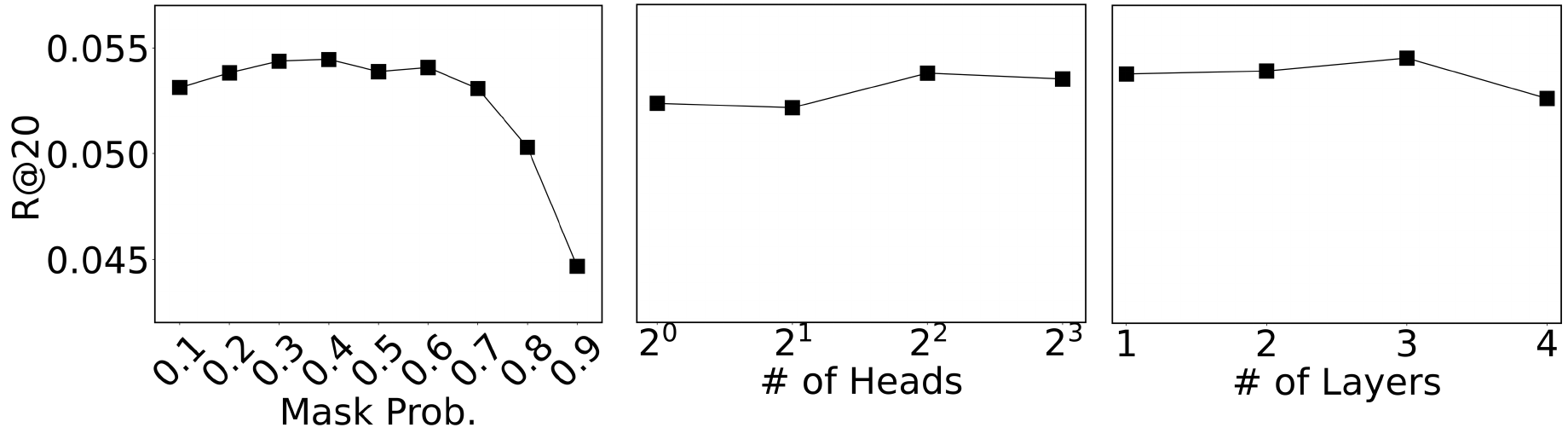}
    \caption{Sensitivity analysis of \proposed.}
    \label{fig:sensitivity}
\end{figure}

\section{Sensitivity Analysis}
\label{subsec:sensitivity}
We provide a sensitivity analysis to guide the hyperparameter selection of \proposed.
In Fig~\ref{fig:sensitivity}, we report the recommendation accuracy (R@20) with varying random masking probabilities, the number of heads, and the number of layers for MDRAU-MT on P1. 
Similar tendencies are observed in the other scenario.
We first observe that the masking probability has a small impact within the range of [0.1, 0.7].
However, the performance drastically drops when the masking probability exceeds 0.7.
If the masking probability is too high, the model cannot obtain sufficient information from other domains, leading to degraded performance.
Also, we observe stable performances with varying numbers of heads and layers of the multi-domain encoder. 
\proposed achieves a slightly improved performance with more number of heads, and a slightly degraded performance when the number of layers exceeds~3.

\end{document}